\documentclass{iopart}

\usepackage{graphicx,epsfig,sidecap}
\usepackage{bm}

\def\be{\begin{equation}}
\def\ee{\end{equation}}

\def\bea{\begin{eqnarray}}
\def\eea{\end{eqnarray}}

\def\Tr{{\rm Tr}\,}

\def\e{\epsilon}

\begin{document}

\title[Entanglement and correlation functions following a local quench: a CFT approach]
{Entanglement and correlation functions following a local quench:
a conformal field theory approach}

\author{Pasquale Calabrese$^1$ and John Cardy$^{2}$}
\address{$^1$Dipartimento di Fisica dell'Universit\`a di Pisa and INFN,
             Pisa, Italy.\\
         $^2$ Oxford University, Rudolf Peierls Centre for
          Theoretical Physics, 1 Keble Road, Oxford, OX1 3NP, United Kingdom
          and All Souls College, Oxford.}

\date{\today}

\begin{abstract}
We show that the dynamics resulting from preparing a
one-dimensional quantum system in the ground state of two
decoupled parts, then joined together and left to evolve unitarily
with a translational invariant Hamiltonian (a local quench), can
be described by means of quantum field theory. In the case when
the corresponding theory is conformal, we study the evolution of
the entanglement entropy for different bi-partitions of the line.
We also consider the behavior of one- and two-point correlation
functions. All our findings may be explained in terms of a
picture, that we believe to be valid more generally, whereby
quasiparticles emitted from the joining point at the initial time
propagate semiclassically through the system.

\end{abstract}

\maketitle

\section{Introduction.}

The understanding of the degree of entanglement in extended
quantum systems has prompted an enormous effort in a field at the
border of condensed matter physics, quantum information theory,
and quantum field theory. It would take too long to mention all
the topics and problems addressed so far: we refer the reader to a
recent thorough review and extensive bibliography \cite{ent-rev}.

Several measures have been introduced to quantify the entanglement
in an extended system \cite{ent-rev}. Among them we consider here
only the entanglement entropy, which is defined as follows.
Suppose a whole system is in a pure quantum state $|\Psi\rangle$,
with density matrix $\rho=|\Psi\rangle\langle\Psi|$, and an
observer A measures only a subset $A$ of a complete set of
commuting observables, while another observer B may measure the
remainder. A's reduced density matrix is $\rho_A={\rm
Tr}_B\,\rho$. The entanglement entropy is just the von Neumann
entropy \be S_A=-{\rm Tr}_A\,\rho_A\log\rho_A \label{eedef} \ee
associated with this reduced density matrix.
For an unentangled product state, $S_A=0$. Conversely, $S_A$
should be a maximum for a maximally entangled state. One of the
most striking features of the entanglement entropy is its
universal behavior displayed at and close to a quantum phase
transition. In fact, close to a quantum critical point, where the
correlation length $\xi$ is much larger than the lattice spacing
$a$, the long-distance behavior of the correlations in the ground
state of a quantum spin chain are effectively described by a 1+1
dimensional quantum field theory. At the critical point, where
$\xi$ diverges, the field theory, if Lorentz invariant, is also a
{\em conformal} field theory (CFT). In the latter case, it has
been shown that, if $A$ is an interval of length $\ell$ in an
infinite chain, $S_A\simeq (c/3)\log\ell$
\cite{Holzhey,Vidal,cc-04}, where $c$ is the central charge of the
corresponding CFT. When the correlation length $\xi$ is large but
finite (more precisely $\xi\ll \ell$), it has been shown that,
increasing $\ell$, $S_A$ saturates \cite{Vidal} to $S_A\simeq
(c/3)\log\xi$ \cite{cc-04,ccd-07}.

Recently the interest in the properties of entanglement has been
extended to understanding its dynamical behavior. The natural
question is how the entanglement propagates through the system
when it is prepared in a state that is not an eigenstate and then
is left to evolve in the absence of any dissipation. The most
common situation studied so far concerns a sudden quench of some
coupling of the model Hamiltonian \cite{cc-05,dmcf-06,eo-06,cdrz-07,hrt-07}. 
The main motivation for the large number of studies of this dynamics
is that it may actually be realized and measured in ultra-cold
atomic systems \cite{expuc}. For a complete list of references on
the subject we refer to our earlier paper \cite{cc-07}.

Here we consider a different situation, known in the literature as
local quench \cite{ep-07}. We will concentrate on the case when a
physical one-dimensional system (e.g. a spin chain) is physically
cut into two parts that are in their own ground state. We then
join the two halves at a given time $t=0$ and study the subsequent
evolution according to a translational invariant hamiltonian.
Here, the entanglement entropy is the most natural quantity to be
studied because the two halves are clearly unentangled for $t<0$
whereas the initial energy differs from that of the ground state
only by a finite amount. This differs from the case of a global
quench, when the energy of the initial state above the ground
state is always extensive, and correspondingly the entanglement
can be extensively larger. We will only consider gapless models
that are described asymptotically by a boundary CFT, since in this
case we can use powerful analytic results
\cite{cardy-84,cardy-05}.

The paper is organized as follows. In Sec.~\ref{secCFT} we briefly
recall the CFT approach to the entanglement entropy. We outline
the setup for the local quench which is then applied to different
situations in the following section \ref{secenta}. Then we
consider the time evolution of correlation functions in
Sec.~\ref{secCF}. In Sec.~\ref{seccur} we describe how this
approach can be generalized to the case with two joining points
and we derive an heuristic solution. Finally we conclude the paper
with a discussion of open problems and possible generalizations in
Sec.~\ref{secC}.

\section{The CFT approach to entanglement entropy and a local quench}
\label{secCFT}

\subsection{Entanglement entropy and CFT}

The entanglement entropy $S_A$ defined by Eq. (\ref{eedef})
can be studied in a
general quantum field theory through the replica trick \cite{cc-04}
\be
S_A=\left.-\frac{\partial}{\partial n} {\rm Tr}\,\rho_A^n\right|_{n=1}\,.
\ee
This is particular useful in a CFT because, when $A$ consists of
disjoint intervals with $N$ boundary points with $B$, ${\rm
Tr}\,\rho_A^n$ transforms under a general conformal transformation
as the $N$ point function of a primary field $\Phi_n$ with conformal
dimension \cite{cc-04}
\be
x_n= \frac{c}{12} \left(n-\frac1n\right)\,,
\ee
where $c$ is the central charge of the underlying CFT.

In particular this implies that the entanglement entropy of a slit
of length $\ell$ in an infinite system is given by \be {\rm
Tr}\,\rho_A^n= c_n \left(\frac\ell{a}\right)^{-2x_n}
\qquad\Rightarrow\qquad S_A=\frac{c}3 \log \frac{\ell}{a}+ c'_1\,,
\ee where $a$ is an UV cutoff (e.g. in a spin chain is the lattice
spacing). The constants $c_n$ are also non-universal and so is the
derivative for $n=1$, $c'_1$, entering in $S_A$. However, in any
particular model, $a$ and $c'_1$ can be fixed unambiguously by
specific requirements (see e.g. \cite{ccd-07}).

Another important result we will use in the following is the
entanglement of the slit $A= [0,\ell]$ in a semi-infinite line
with specific boundary conditions at $r=0$, that is given by
\cite{cc-04} \be {\rm Tr}\,\rho_A^n= \tilde{c}_n
\left(\frac{2\ell}{a}\right)^{-x_n} \qquad\Rightarrow\qquad
S_A=\frac{c}6 \log \frac{2\ell}{a}+ \tilde{c}'_1\,. \label{SAboun}
\ee In analogy to the $c_n$, the constants $\tilde{c}_n$ are not
universal and depend on the boundary condition at $r=0$. However
comparing the finite temperature result of Ref. \cite{cc-04} with
the standard thermodynamic entropy for systems with boundaries
\cite{al-91} we have \cite{cc-04,zbfs-06,lsca-06} \be
\tilde{c}'_1-c'_1/2=\log g\,, \ee where $\log g$ is the boundary
entropy first introduced by Affleck and Ludwig \cite{al-91}. We
recall that $g$ is universal and and depends only on the boundary
CFT.

\subsection{CFT for global quenches}

CFT is also able to predict the time-dependence of the
entanglement entropy after a global quench \cite{cc-05}. We
briefly summarize here these results, because some of them will be
useful in understanding the local case. The density matrix has the
path integral representation \cite{cc-05} \be\fl
\langle\psi''(r'')|\rho(t)|\psi'(r')\rangle=Z_1^{-1}
\langle\psi''(r'')|e^{-itH-\e H}|\psi_0(r)\rangle
\langle\psi_0(r)|e^{+itH-\e H}|\psi'(r')\rangle\,, \label{dm0} \ee
where we included damping factors $e^{-\e H}$ in such a way as to
make the path integral absolutely convergent. Each of the factors
may be represented by an analytically continued path integral in
imaginary time: the first one over fields $\psi(r,\tau)$ which
take the boundary values $\psi_0(r)$ on $\tau=-\e-it$  and
$\psi''(r)$ on $\tau=0$, and the second with $\psi(r,\tau)$ taking
the values $\psi'(r)$ on $\tau=0$ and $\psi_0(r)$ on $\tau=\e-it$.
When $H$ is critical and the field theory is a CFT, under the
renormalization group any {\it translationally} invariant boundary
condition is supposed to flow into a boundary fixed point,
satisfying conformal boundary conditions. Thus we may assume that
the state $|\psi_0\rangle$ corresponds to such boundary conditions
on sufficiently long length scales. Thus, for real  $\tau$, the
strip geometry described above may be obtained from the upper
half-plane (where the entanglement entropy is known from
\cite{cc-04}) by the conformal mapping $w=(2\e/\pi)\log z$. In
this way one easily get the result for imaginary time. The real
time evolution is obtained by continuing $\tau=\e+it$. The final
result is \cite{cc-05} \be S_A(t) \simeq \cases{ \frac{\pi
ct}{6\e}       & $t<\ell/2$ \,,\cr \frac{\pi c\,\ell}{12\e} &
$t>\ell/2$ \,. } \ee From this we note that $\e$ enters in the
calculation in an essential way. A careful analysis
\cite{cc-06,cc-07} shows that in fact it corresponds to the
correlation length in the initial state and so it is not only a
useful tool, but a physical important parameter.

The fact that $S_A(t)$ increases linearly until it saturates at
$t^*=\ell/2$ has a simple interpretation in terms of
quasiparticles excitations emitted from the initial state at $t=0$
and freely propagating with velocity $v=1$. This phenomenon is
very general and holds even for non-critical systems, and it has
been confirmed by exactly integrable dynamics and numerics (see
e.g. \cite{cc-05,dmcf-06,ep-07}) However, since in lattice models
there are particles moving slower than $v$, after $t^*$ the
entropy does not saturate abruptly, but is a slowly increasing
function of the time. The same picture is valid for the
correlation functions. Firstly incoherent quasi-particles arriving
a given point from well-separated sources cause relaxation of
(most) local observables towards their ground-state expectation
values. Secondly, entangled quasiparticles (emitted from an
initially correlated region) arriving at the same time $t$ induce
correlations between local observables. In the case where they
travel at a unique speed $v$, therefore, there is a sharp
``horizon'' effect: the connected correlations do not change
significantly from their initial values until time $t\sim |r|/2v$.
After this they rapidly saturate to time-independent values. For
large separations, these decay exponentially differently from the
power law dependence in the ground state. Also for correlation
functions, this picture has been shown to be valid in several
models (see e.g. \cite{cc-06,cc-07,c-06,l-07,cdeo-06,swvc-07}).

It is clear that a very similar interpretation should apply also for
local quenches, but with the fundamental difference that now the
quasi-particle excitations are emitted only from the point where the
quench happened and not from everywhere.

\subsection{CFT setup for local quenches}

\begin{SCfigure}
\includegraphics[width=7cm]{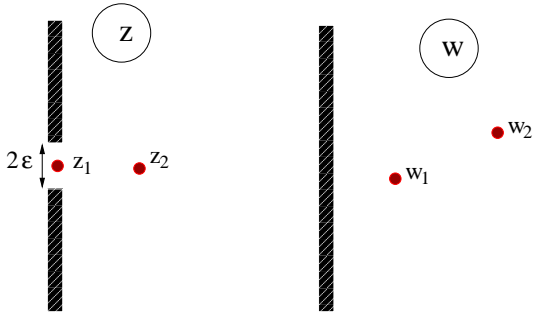} \caption{Space-time region for the
density matrix (left) mapped to the half-plane (right) by means of
Eq. (\ref{mapp1}). $z_1=i\tau$ and $z_2=i\tau+\ell$ during the
computation and in the end $\tau\to i t$.}
\label{map1}
\end{SCfigure}

Suppose we physically cut a spin chain at the boundaries between
two subsystems A and B, and prepare a state where the individual
pieces are in their respective ground states. In this state the
two subsystems are completely unentangled, and its energy differs
from that of the ground state by only a finite amount. Let us join
up the pieces at time $-t$ and watch the system evolve up to
$t=0$.

The procedure for the global quenches does not apply because the
initial state is not translational invariant and will not flow
under the renormalization group toward a conformally invariant
boundary state. One may try to handle the problem introducing
proper boundary condition changing operators (see e.g.
\cite{al-95}). However we prefer to take a different approach. In
fact, we can simply represent the corresponding density matrix in
terms of path integral on a modified world-sheet. The physical cut
corresponds to having a slit parallel to the (imaginary) time
axis, starting from $-\infty$ up to $\tau_1=-\e-i t$ (the time
when the two pieces have been joined), and analogously the other
term of the density matrix, like in Eq. (\ref{dm0}), gives a slit
from $\tau_2=\e-i t$ to $+\infty$. Again we introduced the
regularization factor $\e$, that we will interpret a posteriori.

For computational simplicity we will consider the translated
geometry with two cuts starting at $\pm i\e$ and operator inserted
at imaginary time $\tau$. This should be considered real during
the course of all the computation, and only at the end can be
analytically continued to $it$. This plane with the two slits is
pictorially represented on the left of Fig. \ref{map1} where
$i\tau$ corresponds to $z_1$. As shown in the same figure, the
$z$-plane is mapped into the half-plane ${\rm Re}\, w>0$ by means
of the conformal mapping \be
w=\frac{z}{\e}+\sqrt{\left(\frac{z}{\e}\right)^2+1} \qquad {\rm
with \,\, inverse} \qquad z=\e \frac{w^2-1}{2w}\,. \label{mapp1}
\ee On the two slits in the $z$ plane (and so on the imaginary
axis in the $w$ one) conformal boundary conditions compatible with
the initial state must be imposed. For example, in the most
natural situation when the boundary ``spins'' are left free, we
require free boundary conditions. Oppositely, when the boundary
spins are supposed to stay in a particular state we require fixed
boundary conditions.

Note that this setting is valid in any dimension when the system
is prepared in two spatially divided halves. In this case one can
try to tackle the problem with the methods of boundary critical
phenomena \cite{dd}, but this will be extremely difficult, if not
impossible, since conformal invariance is far less powerful.

\section{Entanglement entropy}
\label{secenta}

\begin{figure}[t]
\epsfig{width=9cm,file=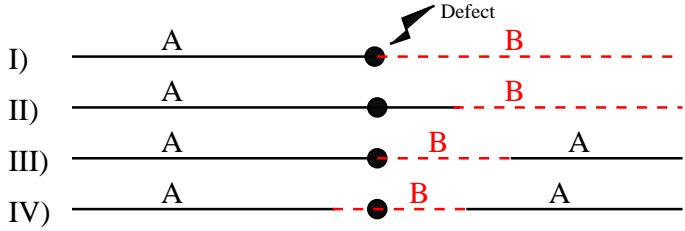} \caption{The four different
bipartitions of the line we consider here.}
\label{confs}
\end{figure}

In this section we consider the time evolution of the entanglement
entropy after the local quench. We consider two half-chains joined
together at the point $r_D=0$ at a given time. The various
subcases we consider correspond to different spatial partitions of
the system among which we calculate the entanglement. We consider
the four different situations depicted in Fig. \ref{confs}.

We start with the more natural division, considering the
entanglement entropy between the two parts in which the system was
divided before the quench (case I). As case II we consider the
entanglement of the region $r>\ell$ with $r<\ell$. This will allow
us to identify simple horizon effects. To highlight the
interaction of the two previous effects we then consider the
entanglement of the part $B=[0,\ell]$ with the rest (case III).
Finally we consider as IV the most general slit
$B=[\ell_1,\ell_2]$. Clearly cases I and III are particular
choices of II and IV respectively and can be derived as simple
limits. However we prefer to present the results in such order
because the physical effects and their interpretation will emerge
in a more natural way.

\subsection{Case I: Entanglement of the two halves}

This is the case when B is the positive real axis and A is the negative real
axis.
$\Tr \rho_A^n$ transforms like a one-point function that in the
$w$ plane is $[2{\rm Re} w_1]^{-x_n}$. Thus in the $z$ plane at
the point $z_1=(0,i\tau)$ we have \be \langle\Phi_n(z_1) \rangle=
\tilde{c}_n \left( \left|\frac{dw}{dz}\right|_{z_1}
\frac{a}{[2{\rm Re} w_1]}\right)^{x_n} \ee that using \be \e w_1=
i\tau+\sqrt{\e^2-\tau^2}\,,\qquad
\left|\frac{dw}{dz}\right|_{z_1}=\frac\e{\sqrt{\e^2-\tau^2}}\,,
\label{z1map} \ee becomes \be
\langle\Phi_n\rangle=\tilde{c}_n\left(\frac{a\e/2}{\e^2-\tau^2}\right)^{x_n}\,.
\label{phin} \ee Continuing this result to real time $\tau\to it$
we obtain \be
\langle\Phi_n(t)\rangle=\tilde{c}_n\left(\frac{a\e/2}{\e^2+t^2}\right)^{x_n}\,.
\ee Using finally the replica trick to find the entanglement
entropy we have \be S_A=- \left.\frac{\partial }{\partial n}\Tr
\rho_A^n\right|_{n=1}= \frac{c}{6}\log \frac{t^2+\e^2}{a\e/2}
+\tilde{c}'_1\,. \label{Shalf} \ee There are two main pieces of
information that we can extract from this result, coming from the
long and the short time behavior. For $t\gg\e$ we have \be
S_A(t\gg\e)=\frac{c}{3}\log \frac{t}a +k_0\,, \label{Slogt} \ee
i.e. the leading long time behavior is only determined by the
central charge of the theory in analogy with the ground state
value for a slit. This could result in a quite powerful tool to
extract the central charge in time-dependent numerical
simulations. This behavior resembles the slow increasing of $S_A$
observed numerically in Ref. \cite{gkss-05} (unfortunately these
data were never fitted, to understand the quantitative predictive
power of our formula). The constant $k_0$ is given by
$k_0=\tilde{c}'_1+(c/6)\log(2a/\e)$.

The behavior for short time allows instead to fix the regulator
$\e$ in terms of the non-universal constant $\tilde{c}'_1$. In
fact we have \be S_A(t=0)= \frac{c}{6}\log \frac{2\e}{a}
+\tilde{c}'_1=0 \qquad \Rightarrow \qquad \e= \frac{a}2
e^{-6\tilde{c}'_1/c}\,. \label{epsfixing} \ee Note that, even if
non-universal, for a given lattice model the constant
$\tilde{c}'_1$ can be fixed only by ground state quantities.
Consequently Eq. (\ref{Shalf}) has no free dynamical parameter.

The parameter $\e$ depends in a specific manner on the boundary
contribution to the entanglement $\tilde{c}'_1$. This is
completely different from what happens for a global quench where
the equivalent regulator is connected to the correlation length in
the initial state \cite{cc-06,cc-07}. Furthermore the order of
magnitude of $\e$ is fixed by the lattice spacing $a$ (i.e. the UV
cutoff) and so considering the limit of times and lengths larger
than $\e$ is equivalent to the standard condition to apply the
field theory for distances larger than $a$.

\subsection{Case II: Decentered defect}

Let us now consider the entanglement of the region $r>\ell$ with
the rest of the system. In this case $\Tr \rho_A^n$ is equivalent
to the one-point function in the plane $z$ at the point
$z_2=\ell+i\tau$ as in Fig. \ref{map1}. Under the conformal
mapping (\ref{map1}) this point goes into \be \e w_2= \ell
+i\tau+\sqrt{\e^2+(\ell+i\tau)^2} \equiv \ell +i\tau+\rho
e^{i\theta} \label{z2map} \ee with \be\fl \rho^2=
\sqrt{(\e^2+\ell^2-\tau^2)^2+4\ell^2 \tau^2}\,,\qquad
\theta=\frac12\arctan\frac{2\ell\tau}{\sqrt{\e^2+\ell^2-\tau^2}}\,,
\label{rthdef} \ee and \be \left|\e\frac{dw}{dz}\right|_{z_2}=
\frac{\sqrt{(\ell+\rho\cos\theta)^2+(\tau+\rho\sin\theta)^2}}\rho\,.
\label{dz2map} \ee Thus $\langle\Phi_n\rangle$ is given by Eq.
(\ref{phin}) with $w_1\to w_2$, resulting in \be
\langle\Phi_n\rangle= \tilde{c}_n\left(
\frac{a\sqrt{(\ell+\rho\cos\theta)^2+(\tau+\rho\sin\theta)^2}}{
2\rho(\ell+\rho\cos\theta)} \right)^{x_n}\,. \label{boh} \ee The
real time evolution for $t,\ell\gg \e$ is obtained by analytically
transforming $\tau\to it$. The calculation can seem very
cumbersome, but it is greatly simplified by the fact that in this
limit we have $\rho^2\to |\ell^2-t^2|$ and $\rho \cos\theta\to
\max[\ell,t]$,  $\rho\sin\theta\to i \min[\ell,t]$. Care must also
be taken when the zero-th order in $\e$ is vanishing. In the end
we have \be
\frac{\sqrt{(\ell+\rho\cos\theta)^2+(\tau+\rho\sin\theta)^2}}{
2\rho(\ell+\rho\cos\theta)}\to \cases{ \frac1{2|\ell|} &
$t<\ell$\,,\cr \frac\e{2(t^2-\ell^2)}&  $t>\ell$\,, } \ee that
leads to \be S_A= \cases{ \frac{c}6 \log \frac{2\ell}a +
\tilde{c}'_1  & $t<\ell$\,,\cr \frac{c}6 \log
\frac{t^2-\ell^2}{a^2} + k_0 & $t>\ell$\,, } \ee with $k_0$ the
same as in Eq. (\ref{Slogt}). The interpretation of this result is
quite direct. Indeed at $t=0$ the joining procedure produces a
quasi-particle excitation at $r=0$ that propagates freely with the
corresponding speed of sound $v_s$ that in the CFT normalization
is $v_s=1$. This excitation takes a time $t=\ell$ to arrive at the
border between $A$ and $B$ and only at that time will start
modifying their entanglement. The following evolution for
$t\gg\ell$ is the same as in Eq. (\ref{Slogt}).

Also the constant value for $t<\ell$ deserves a comment: it is
exactly the value known from CFT for the slit in the half-line Eq.
(\ref{SAboun}). This is a non-trivial consistency check. Note that
a finite $\e$ smooths the crossover between the two regimes and
makes the entanglement entropy a continuous function of the time.

\subsection{Case III: The slit with the defect at the border}

Let us consider again the same physical situation as before, but
we now calculate the entanglement entropy of $A=[0,\ell]$ and $B$
the remainder. For $t<0$ the real negative axis is decoupled from
the rest and does not contribute to the entanglement entropy, that
is just the one of a slit in half-chain, i.e. the initial entropy
is given by Eq. (\ref{SAboun}).

The entanglement entropy is obtained from the replica trick
considering the scaling of a two-point function between the
endpoints of the slit. In the $z$ plane these two points are
$z_1=i\tau$ and $z_2= \ell+i\tau$ (we adopt the same notation as
before to make direct the use of previous formulas).
As usual, the plane with cuts
is mapped into the half-plane ${\rm Re}\, w>0$ by the transformation
(\ref{mapp1}). On the $w$ plane we have \cite{cc-04}
\be\fl
\langle\Phi_n(w_1)\Phi_{-n}(w_2)\rangle=\tilde{c}_n^2
\left(\frac{a^2|w_1+\bar{w}_2||w_2+\bar{w}_1|}{
|w_1-w_2| |\bar{w}_2-\bar{w}_1|  |w_1+\bar{w}_1||w_2+\bar{w}_2|}
\right)^{x_n}\,,
\label{2pt}
\ee
so the mapping to the original plane $z$ is
\be\fl
\langle\Phi_n(z_1)\Phi_{-n}(z_2)\rangle=\tilde{c}_n^2
\left(\left|\frac{dw}{dz}\right|_{z_1}
\left|\frac{dw}{dz}\right|_{z_2}
\frac{a^2|w_1+\bar{w}_2||w_2+\bar{w}_1|}{
|w_1-w_2| |\bar{w}_2-\bar{w}_1|  |w_1+\bar{w}_1||w_2+\bar{w}_2|}
\right)^{x_n}
\ee
where the various terms are functions of $z_i$ through
Eqs. (\ref{z1map}), (\ref{z2map}), (\ref{rthdef}), (\ref{dz2map}), and
\bea
\e^2|w_1-w_2|^2&=&
(\ell+\rho\cos\theta-\sqrt{\e^2-\tau^2})^2+\rho^2\sin^2\theta\,,\\
\e^2|w_1+\bar{w}_2|^2&=&
(\ell+\rho\cos\theta+\sqrt{\e^2-\tau^2})^2+\rho^2\sin^2\theta\,.
\eea
Putting everything together we get
\bea
\fl \langle\Phi_n(z_1=i\tau)\Phi_{-n}(z_2=i\tau+\ell)\rangle=\nonumber \\
\fl \tilde{c}_n^2 \left(
\frac{a^2\e}{\e^2-\tau^2}
\frac{(\ell+\rho\cos\theta+\sqrt{\e^2-\tau^2})^2+\rho^2\sin^2\theta}{
(\ell+\rho\cos\theta-\sqrt{\e^2-\tau^2})^2+\rho^2\sin^2\theta}
\frac{\sqrt{(\ell+\rho\cos\theta)^2+(\tau+\rho\sin\theta)^2}}{
4\rho(\ell+\rho\cos\theta)}
\right)^{x_n}\,
\eea
Again this simplifies considering the limit $t,\ell\gg\e$ as
explained after Eq. (\ref{boh}). We finally obtain
\be
{\rm Tr} \rho_A^n=
 \cases{
 \tilde{c}_n^2
 \left(\frac{a^2}{t^2}\frac{\ell+t}{\ell-t}\frac\e{4\ell}
 \right)^{x_n}&       $t<\ell$\,,\cr
 \tilde{c}_n^2\left(\frac{a^2}{\ell^2}\right)^{x_n} &
 $t>\ell$\,.
}
\ee
Note that for $t>\ell$ the calculation is slightly more cumbersome
because we should take into account the first order in $\e$ of some
expressions.

\begin{figure}[b]
\caption{$S_A(t)$ for a general slit $B=[\ell_1,\ell_2]$.
The plots are done for finite $\e=10^{-3}$, and they are indistinguishable from
the asymptotic result. In all the plots $\ell_1-\ell_2=20$.
Left: $\ell_2<0$, for several $\delta= \ell_1+\ell_2$.
Right: $\ell_2>0$, for several $\ell_1$.}
\vspace{6mm}
\centerline{\epsfig{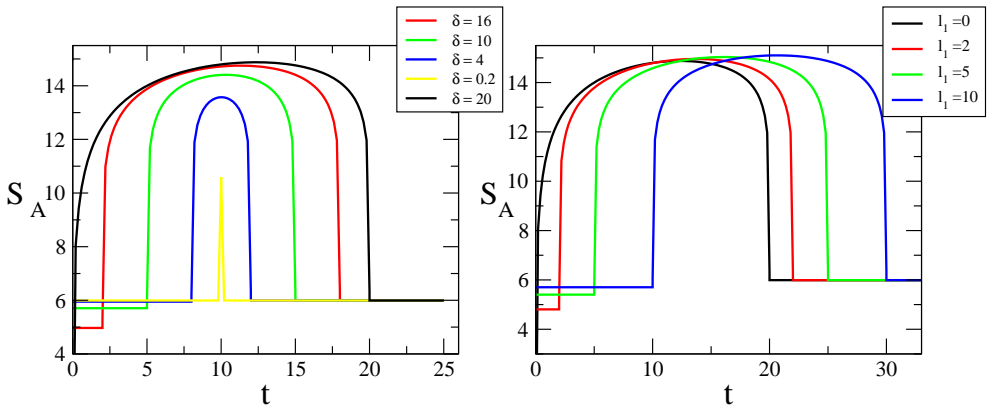}}
\label{Figslit}
\end{figure}

Using finally the replica trick we get for the entanglement entropy
\be
S_A=\cases{
\frac{c}3 \ln \frac{t}a+\frac{c}6 \ln\frac\ell\e
+\frac{c}6 \ln 4\frac{\ell-t}{\ell+t}+2\tilde{c}'_1&
 $t<\ell$\,, \label{plateau}\cr
\frac{c}3 \ln \frac{\ell}a +2\tilde{c}'_1& $t>\ell$\,.
}
\ee
The crossover time $t^*=\ell$ is again in agreement with the quasi-particles
interpretation.

There are several interesting features of this result. For very
short time $t\ll \ell$ it reduces to the $\ell=\infty$ case Eq.
(\ref{Slogt}) as it should. The leading term for $t>\ell$ is just
the ground state value for a slit in an infinite line. However the
subleading term is not the same, in fact we have \be
S_A(t>\ell)-S_A({\rm eq})=2\tilde{c}'_1-c'_1=\log g\,, \ee where
we also use the value of $\e$ in Eq. (\ref{epsfixing}). This is a
signal that for long time the system still remembers something of
the initial configuration as a boundary term that is unable to
``dissipate''. Since the extra energy never dissipates under
unitary evolution, there is no reason for the constant terms to be
the same.

Another interesting feature is the behavior for $t<\ell$. This is
very similar to the form proposed in Ref.~\cite{ep-07} to fit the
numerical data, i.e. \be S_A=\frac{c_0}3\log \ell
+\frac{c_1}{3}\log(t/\ell)+\frac{c_2}3\log(1-t/\ell)+k'\,. \ee
Only the term in $t+\ell$ was missing in Ref.~\cite{ep-07}.
However this behaves smoothly for $0<t<\ell$ and its effect can be
well approximate by a constant factor in $k'$. It will be
interesting to check whether and how the use of this term in
$\ell+t$ changes the quality of the fit. Furthermore the fit of
Ref. \cite{ep-07} is also a strong confirmation of our result,
indeed they found $c_0\simeq 1+c_2$, $c_1\simeq 1$, and $c_2\simeq
1/2$ (plot 7 in Ref. \cite{ep-07} with $t'=0$) that are exactly
our predictions for $c=1$ (the known central charge of the XX
model).

Furthermore Eq. (\ref{plateau}) display a large plateau for $0.2 <t/\ell< 0.8$
as shown in Fig. \ref{Figslit} (the black line in both the plots)
and already noticed in Ref. \cite{ep-07}.
This plateau
can be studied considering the value at the maximum $t/\ell=(\sqrt5-1)/2$
where we have
\be
S_A({\rm plateau})=\frac{c}2 \log\frac{\ell}a+k_1= \frac32 S_A({\rm
eq})+k_2\,,
\ee
with the constants $k_p$ simply related to $a$, $\e$, and
$\tilde{c}'_1$.

Finally we can study the result for $t=0$ when $\e\ll\ell$: \be
S_A(t=0)=\frac{c}6 \log \frac{4\e\ell}{a^2}+2\tilde{c}'_1=
\frac{c}6 \log \frac{2\ell}{a}+\tilde{c}'_1\,, \ee where in the
last equality we used the value of $\e$ given by Eq.
(\ref{epsfixing}). This result is exactly the initial value for a
slit of length $\ell$ in the half-line. Again this is an highly
non-trivial consistency check of all the method. A final curiosity
is that \be 2S_A(t=0)-S_A(t>\ell)=\frac{c}3\log 2\,, \ee is
independent of the details of the theory.

\subsection{Case IV: The general slit}
\label{genslit}

Let us now consider the most general case of a slit
$A=[\ell_2,\ell_1]$. In the $z$ plane we have to calculate the
two-point function between $z_1=\ell_1+i\tau$ and
$z_2=\ell_2+i\tau$. Under the mapping (\ref{map1}) these correspond
to $w_p=\ell_p+i\tau+\rho_p e^{i\theta_p}$ with $\rho_p$ and
$\theta_p$ given by Eq. (\ref{rthdef}) with $\ell\to\ell_p$ and
$p=1,2$. In imaginary time, after a long, but straightforward
algebra we have
\bea
\fl \left(\frac{\langle\Phi_n(z_1=i\tau+\ell_1)\Phi_{-n}(z_2=i\tau+\ell_2)
\rangle}{\tilde{c}_n^2}\right)^{1/x_n}=\nonumber\\
\fl ~~~~~~\frac{ \sqrt{(\ell_1 + \rho_1 \cos \theta_1)^2 + (\tau + \rho_1 \sin
\theta_1)^2} \sqrt{(\ell_2 + \rho_2 \cos \theta_2)^2 + (\tau +
\rho_2 \sin \theta_2)^2}}{ 4\rho_1 \rho_2(\ell_1+ \rho_1
\cos\theta_1)(\ell_2 + \rho_2 \cos\theta_2)} \times\nonumber \\
\fl ~~~~~~~~\times
\frac{(\ell_1 + \ell_2+\rho_1 \cos\theta_1 +\rho_2 \cos\theta_2)^2 +
(\rho_1 \sin\theta_1 - \rho_2 \sin\theta_2)^2 }{(\ell_1
-\ell_2+\rho_1 \cos\theta_1 - \rho_2 \cos\theta_2)^2 + (\rho_1
\sin\theta_1 - \rho_2 \sin\theta_2)^2}\,.
\eea

The analytic continuation can be simplified for
$t,\ell_1,\ell_2\gg\e$.
For simplicity in the notation, but without loss of generality, we indicate as
$\ell_1$ the larger in absolute value of the two $\ell_p$ and we assume
$\ell_1>0$. $\ell_2$ can be either positive or negative.
For $t<|\ell_2|$ we have a different result if the two points are on the
same side or on different sides of the defect, namely:
\be\fl
S_A(t<|\ell_2|)=
\cases{
\frac{c}6 \log \frac{4 \ell_1 |\ell_2|}{a^2} +2 \tilde{c}'_1 &
$\ell_2<0$\,, \cr
 \frac{c}6 \log \frac{(\ell_1-\ell_2)^2}{(\ell_1+\ell_2)^2} \frac{4 \ell_1\ell_2}{a^2}
+2 \tilde{c}'_1 & $\ell_2>0$\,. } \ee The first is just the sum of
the entanglement entropies of two slits of length $\ell_1$ and
$|\ell_2|$ in the half-line starting at the origin. In fact, in
the initial state the two parts are unentangled and the total
entropy is the sum of the two. Also the second result is just the
ground state value for the slit $[\ell_2,\ell_1]$ in the half-line
\cite{cc-04}.

The entanglement between $A$ and $B$ becomes sensitive to the
joining of the system when the quasiparticles emitted from the
origin arrive at $\ell_2$. The behavior for
 $|\ell_2|<t< \ell_1$ does not depend on the
relative sign of the $\ell_p$: \be
S_A(|\ell_2|<t<\ell_1)=\frac{c}6 \log \frac{(\ell_1 -
\ell_2)(\ell_1 - t)}{ (\ell_1 + \ell_2)(\ell_1 +t)} \frac{4\ell_1
(t^2 - \ell_2^2)}{\e a^2} +2 \tilde{c}'_1\,. \ee Finally for
$t>\ell_1$ we have almost the ground state value: \be
S_A(t>\ell_1)=\frac{c}3 \ln \frac{\ell_1-\ell_2}a
+2\tilde{c}'_1\,, \ee that is exactly the same as for case III.

The resulting $S_A$ for different values of $\ell_p$ is plotted in
Fig. \ref{Figslit}.
For $|\ell_2|<t<\ell_1$ there is again a large plateau
whose actual extension and value depend on both $\ell_p$.
The analytical value of the
plateau is quite complicated and not really illuminating (it
contains cubic roots, as can be realized solving the equation for
the maximum).
The most important features are that it decreases
when the defect penetrates deeply in the slit (left part of
Fig. \ref{Figslit}) and it stays almost constant when the slit moves far from
the defect (right part of the figure).

Finally one can easily read off from our formula the result for a
defect exactly at the center of the slit: the entanglement entropy
is independent of the time. This is different from what found
numerically in a lattice model \cite{ep-07} and will be discussed
(among the other things) in the next subsection.

\subsection{Comparison with numerical works}
\label{compa}

As far as we are aware the entanglement entropy after a local
quench has been discussed in only two papers \cite{ep-07,gkss-05}.
There are a few qualitative differences between these results and
the asymptotic ones of the CFT that are easily understood in terms
of the general scenario we draw for quasiparticles emitted from
the origin. These differences are attributable to slow
quasiparticles, that exist as consequence of a non-linear
dispersion relation $v_k =\partial_k E_k$, emitted from the
joining point at $t=0$.

The first difference is that the lattice numerics present, on top
of a smooth curve, very fast oscillations. These oscillations have
been discussed in the context of global quenches
\cite{cc-06,cc-07}, and are due to the modes at the zone boundary
$|ka|=\pi$ that have $v_k=0$. However, they are only corrections
to the asymptotic result for $t,\ell\gg a$, since their amplitude
remains constant while the CFT result diverges. Furthermore for
any specific model they can be easily predicted in a quantitative
way.

In Ref. \cite{ep-07} different configurations of slits were
considered for electron hopping on a chain (XX model in spin
language). The sharp horizon effect is present even in the
numerics, since there cannot be (by definition) quasiparticles
faster than the CFT ones. We already discussed that the numerical
results for a defect at the boundary of the slit is in {\it
quantitative} agreement with the CFT in the region $t<\ell$.
However for $t>\ell$ the numerics show an asymptotic value that
appears to be exactly that of the ground state. Accordingly to our
analysis this is possible only when the boundary entropy $\log g$
vanishes.

The most relevant qualitative difference is that for $t$ greater
than the maximum length the asymptotic result is not constant but
it is slowly decaying (like $\log (t) /t$) smoothing the sharp
transition from the plateau. This, for example, changes completely
the behavior in the case of central defect. We expect that this
phenomenon can be ascribed to the slow quasiparticles that, after
the quickest ones have arrived entangling the two parts, then
disentangle them for very long times. This is the most plausible
explanation, but we do not have an argument to put it on a more
quantitative level.

\section{Correlation functions}
\label{secCF}

In this section we derive the time-dependence of correlation
functions after the local quench. We find that the one-point
function whose functional form is completely fixed by conformal
invariance. The two-point functions instead depend on the
particular (boundary) CFT. We will discuss all the details in the
simplest case of a gaussian theory and then show how from general
CFT arguments we can obtain part of the asymptotic behavior. Some
of these results may have been previously derived in the context
of quantum impurity problems from a different point of view (see
e.g. \cite{s-98}).

\subsection{One-point function}

The one-point function of a primary field in the half-plane
 ${\rm Re}\, w>0$ is
\be
 \langle \Phi(w) \rangle= \frac{A^\Phi_b}{[2 {\rm Re}\, w]^{x_\Phi}}\,,
\label{onepoint}
\ee
where $x_\Phi$ is the scaling dimension of the field and $A^\Phi_b$
is a non-universal amplitude that can be fixed in terms of the
normalization of the two-point function of the same operator. We
also fix the lattice spacing $a$ to $1$. $A^\Phi_b$ is known for the
simplest universality classes as the Gaussian theory and the Ising
model \cite{cl-91}.

With the mapping (\ref{mapp1}) we can get the one-point
correlation, that obviously assumes the same form as Eq.
(\ref{boh}). At the point $r$, after continuing to real time we
have \be \langle \Phi(r,t) \rangle= \cases{ A^\Phi_b (2
r)^{-x_\Phi} &\qquad $t<r$\,, \cr A^\Phi_b \left(\frac{\e}{2(t^2-
r^2)}\right)^{x_\Phi} &\qquad $t>r$\,. } \ee Thus for short times
the correlation takes its initial value, until the effect of the
joining arrives at time $t=r$ when it decays for $t\gg r$ like
$t^{-2x_\Phi}$ (note that this exponent is twice the boundary
one).


\subsection{A two-point function: the gaussian model}

For the gaussian model the two-point function of a primary field
$\Phi$ in the half-plane is given by Eq. (\ref{2pt}) with
$(\ell_2,\ell_1)\to(r_2,r_1)$ and $c_n\to (A_b^\Phi)^2=1$
\cite{cardy-84}. As a consequence we can obtain its scaling as a
byproduct of the result for the entanglement entropy for a general
slit in Sec. \ref{genslit}.

For simplicity in the notation we assume here and in the following section
$r_1$ to be positive and to be larger than the absolute
value of $r_2$ that can be either positive or negative.
Thus we have that for $t<|r_2|$ the two-point function keeps its initial value
that has a different form depending on the relative signs of $r_p$
given by $|4r_1 r_2|^{-x_\Phi}$ (different sides of the defect) and
$|4r_1 r_2 (r_1-r_2)^2/(r_1+r_2)^2|^{-x_\Phi}$ (same side).

For $t>r_1$, it reaches the ground state value
$|r_1-r_2|^{-2x_\Phi}$. An interesting and non-trivial behavior is
displayed for $|r_2|<t<r_1$ when \be \langle \Phi(r_1,t)
\Phi(r_2,t)\rangle=\left[ \frac{(r_1 + r_2)(r_2+t)}{(r_1 - r_2)
(r_1 -t)} \frac\e{4r_1 (t^2 - r_2^2)}\right]^{x_\Phi}\,. \ee

\subsection{The general two-point function}

There are some features of the previous result that are expected
to be valid in general, not only for a Gaussian theory, as for
example the the horizon effect and the final equilibrated value.
The natural expectation is that these results are obtainable with
CFT and in fact we will show that this is the case.

The two-point function in the half-plane can be always written as
\cite{cardy-84}
\be\fl
\langle\Phi(w_1) \Phi(w_2) \rangle=
 \left(\frac{|w_1+\bar{w}_2||w_2+\bar{w}_1|}{
|w_1-w_2| |\bar{w}_2-\bar{w}_1|  |w_1+\bar{w}_1||w_2+\bar{w}_2|}
\right)^{x_\Phi} F(\eta)\,,
\label{2ptgen}
\ee
where $\eta$ is the four-point ratio
\be
\eta=\frac{|w_1+\bar{w}_1||w_2+\bar{w}_2|}{|w_1+\bar{w}_2||w_2+\bar{w}_1|}\,,
\ee
and the function $F(\eta)$ depends explicitly on the considered
(boundary) model. It is known for the simplest models as e.g. the
Gaussian ($F(\eta)=1$) and the Ising universality class (see below).

Thus the behavior of the two-point function depends mainly on the
value on the ratio $\eta$ that we now study. Using the
analytic structure of the previous section we have for $t<|r_2|$ a
different behavior if the two points are on the same or on different
sides of the defect:
\be
\eta(t<|r_2|)=\cases{
 \frac{4r_1 r_2}{(r_1 + r_2)^2} & $r_2>0$\,,\cr
 \frac{\e^2 r_1 |r_2|}{(r_1^2-t^2)(r_2^2-t^2)} & $r_2<0$\,.
}
\ee
For intermediate times we have
\be
\eta(|r_2|<t<r_1)=\frac{2 r_1 (r_2 + t)}{(r_1 + r_2)(r_1 + t)}\,,
\ee
while for larger times $\eta$ is $1$ constantly.

From the previous subsection we already know how the first part of
Eq. (\ref{2ptgen}) transforms under the conformal mapping (\ref{mapp1}).
Thus we need only to map $F(\eta)$ that, in
general, is an unknown function. However we know its behavior in two
special circumstances.
Indeed when $\eta\sim1$ the two points are deep in the bulk, meaning
$F(1)=1$. Instead for $\eta\ll 1$, from the short-distance
expansion, we have
\be
F(\eta)\simeq (A^\Phi_b)^2\eta^{x_b},
\ee
where $x_b$ is the boundary scaling dimension of the leading
boundary operator to which $\Phi$ couples and $A^\Phi_b$ is the
bulk-boundary operator product expansion coefficient that equals the
one introduced in Eq. (\ref{onepoint}) [see e.g. \cite{cl-91}].

From this we easily understand the behavior for $t>r_1$. Since
$\eta\to1$ so $F(1)=1$, the two-point function is just the one in
the ground state \be \langle \Phi(r_1<t)
\Phi(|r_2|<t)\rangle=\frac1{|r_1-r_2|^{2 x_\Phi}}\,. \ee Although
this result could have been expected, it is important to have
recovered it only from CFT arguments.

The behavior for $|r_2| <t<r_1 $ is instead more complicated:  combining the
result of the previuos section with the value of $\eta$ we have
\be\fl
\langle \Phi(r_1,t) \Phi(r_2,t)\rangle=\left[
\frac{(r_1 + r_2)(r_2+t)}{(r_1 - r_2) (r_1 -t)} \frac\e{4r_1 (t^2 -
r_2^2)}\right]^{x_\Phi}
F\left( \frac{2 r_1 (r_2 + t)}{(r_1 + r_2)(r_1 + t)}\right) \,.
\ee
For example when $\Phi$ is the order parameter in the
Ising universality class we have \cite{cardy-84}
\be
F(\eta)=\frac{\sqrt{1+\eta^{1/2}}\pm \sqrt{1-\eta^{1/2}}}{\sqrt{2}}\,,
\label{defF}
\ee
and $x_\Phi=1/8$. The sign $\pm$ depends on the boundary conditions. $+$
corresponds to fixed boundary conditions and and $-$ to free ones.
This completely fixes the behavior of the two-point function in the
intermediate regime. It would be very interesting to check this prediction.

The time evolution for $t<|r_2|$ depends on the sign of $r_2$. For
$r_2>0$ we have \be
 \langle \Phi(r_1,t) \Phi(r_2,t)\rangle=\left[
\frac1{4 r_1 r_2} \frac{(r_1+r_2)^2}{(r_1-r_2)^2}\right]^{x_\Phi}
F\left(\frac{4r_1 r_2}{(r_1 + r_2)^2}\right)\,,
\ee
i.e. it keeps its inital value until $t=r_2$, that is nothing but correct
boundary value at zero time.

More complicated is the behavior for $r_2<0$, in fact we get
\begin{eqnarray}
 \langle \Phi(r_1,t) \Phi(r_2,t)\rangle&=&
\frac1{|4 r_1 r_2|^{x_\Phi}} F\left(\frac{\e^2 r_1
|r_2|}{(r_1^2-t^2)(r_2^2-t^2)}\right)\nonumber\\&\simeq&
\frac{(A_b^{\Phi})^2}{|4 r_1 r_2|^{x_\Phi}} \left(\frac{\e^2 r_1
|r_2|}{(r_1^2-t^2)(r_2^2-t^2)}\right)^{x_b}\,,
\end{eqnarray}
where in the last approximation we used that $\e\ll t,r_1,r_2$ and
the behavior of $F(\eta)$ for $\eta\ll1$. In particular we note
that this is time independent only when $x_b=0$, which corresponds
to the case of a non-zero 1-point function. Similar anomalous
time-behavior was found when $x_b>0$ in Ref.~\cite{cc-05}.

\section{Decoupled finite interval}
\label{seccur}

A natural question arising is how the results we just derived
change when we introduce more than one defect in the line. It is
straightforward to have a path integral for the density matrix: we
only need to have pairs of slits for $-\infty$ to $-i\e$ and from
$i\e$ to $+i\infty$ everywhere there is a defect. However it
becomes prohibitively difficult to treat this case analytically.
In order to begin to understand the case when a finite interval
interval is initially decoupled, we consider the case when it lies
at the end of a half-line.

\begin{figure}[t]
\centerline{\epsfig{width=13cm,file=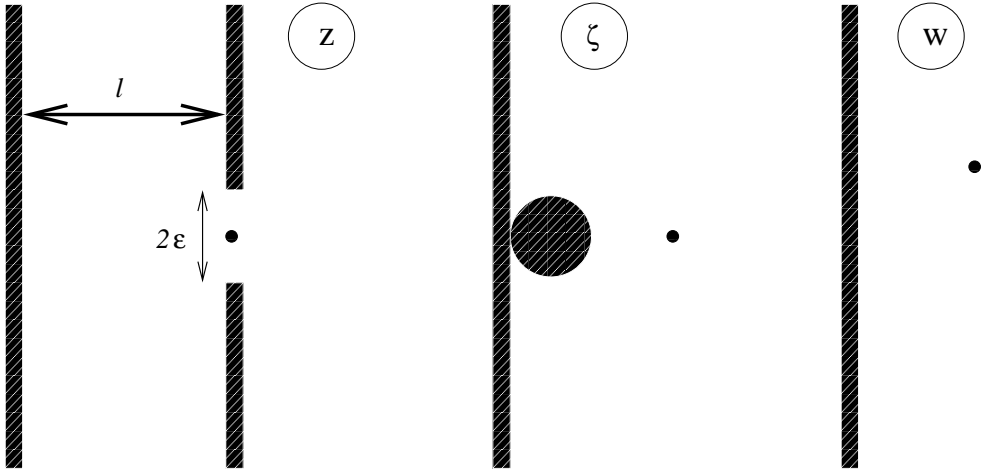}}
 \caption{Left:
Space-time region for the density matrix of a semi-infinite system.
Center: Mapping of the $z$-plane under the transformation Eq.
(\ref{mapp1}). Right: Final map to the half-plane of the $\zeta$
plane given by Eq. (\ref{mapp2}) when $\ell\gg\e$.}
\label{map2}
\end{figure}

So, let us consider a semi-infinite chain in which the A
subsystem is the finite segment $(-\ell,0)$ and the B is the
complement $(0,\infty)$ and with the initial defect at $r_D=0$.
The space-time geometry describing this
situation is like the one just considered, with a wall at $-\ell+iy$
($y$ real) that represents the boundary condition. This is depicted
in the left panel of Fig. \ref{map2}.

In these circumstances the inverse conformal mapping between the
$z$ plane and the half-plane can be worked out using the
Schwarz-Christoffel formula. After long algebra one obtains \be
z(w)=i\left(\frac{\ell}{\pi}\log (iw)+b
\frac{-iw-1}{-iw+1}\right)\,, \label{exmap} \ee with the parameter
$b$ related to $\ell$ and $\e$ in a non-algebraic way. (A slit in
the full line is closely related to this transformation, the last
piece is replaced by $(w^2-1)/(w^2+1)$.) Unfortunately the mapping
(\ref{exmap}) is not analytically invertible and its exact use is
limited to numerical calculations that do not help us, since we
need to perform an analytical continuation. We will develop in the
remaining part of the section an approximate solution for $\ell\gg
\e$, that however is not always justified: this limit is allowed
only {\it after} the analytical continuation to real time.

The limit $\ell\gg \e$ before the analytical continuation simplifies
the calculation because under the conformal transformation
(\ref{mapp1}) the boundary at $x=-\ell$ is approximately a disk
tangent to the imaginary axis at the origin and with radius $R=\e/4\ell\ll
1$, as depicted in the central part of Fig. \ref{map2}. This is
simply checked applying the inverse transformation (\ref{mapp1}) to
the disk $\zeta=R(1+e^{i\theta})$ with $\theta\in [0,2\pi]$. In fact
we have
\be\fl
z=\e\frac{\zeta^2-1}{2\zeta}=\e
\frac{R^2(1+e^{i\theta})^2(1+e^{-i\theta})-(1+e^{-i\theta})}{4R (1+\cos\theta)}
=-\frac{\e}{4R}- i\frac{\e}{4R} \tan \theta/2 +O(R)\,.
\ee
The half-plane minus the disk $R(1+e^{i\theta})$ is mapped in the
half-plane ${\rm Re}\,w>0$ by the transformation
\be
w=-i\exp(2\pi i R/\zeta)\,.
\label{mapp2}
\ee
Let us note that when $\e\ll\ell$, the parameter $b$ is given by
$b=\pi\e^2/(8\ell)+O(\e^4/\ell^2)$.

Combining the two transformations we can map the space-time region
on the left of Fig. \ref{map2} into the half-plane (only in the
limit $\ell\gg \e$) by means of \be w=-i\exp\left[ \frac{\pi i
\e}{2\ell}\left(\sqrt{z^2/\e^2+1}-z/\e \right)\right]\,, \ee with
inverse \be z(w)=i\frac{\ell}{\pi}\log(iw)+ i\frac{\e^2}{\ell}
\frac{\pi}{4} \frac{1}{\log(iw)}\,. \ee 
Note that the residue of the pole at $w=-i$ is the same as the exact 
one in Eq. (\ref{exmap}).
Furthermore, calculating the the difference between the
approximate and the exact solution, one easily checks that it is
of the order of $\e^4/\ell^2$ everywhere in the complex plane.

We then have that $w(i\tau)$ can be written as
\be\fl
w(i\tau)=-ie^{\pi\tau/2\ell}\left[\cos(\sqrt{1-\tau^2/\e^2}\pi\e/2\ell)+
i\sin(\sqrt{1-\tau^2/\e^2}\pi\e/2\ell)\right]\,.
\ee
Thus
\be
\langle\Phi_n(i\tau)\rangle= \tilde{c}_n
\left[\frac{\pi\e}{4\ell}\frac{1}{\sqrt{\e^2-\tau^2}}
\frac{1}{\sin(\pi \sqrt{\e^2-\tau^2}/2\ell)} \right]^{x_n}\,.
\ee
Note that for $\ell\to \infty$ it reduces to the previous result, as it should.
Furthermore, it is clear that it is well defined only for $\ell\gg\e$.

Continuing to real time $\tau\to it$, and for $t\gg\e$ we have \be
\langle\Phi_n(i\tau)\rangle=\tilde{c}_n
\left[\frac{\pi\e}{4\ell}\frac{1}{t\sin(\pi
t/2\ell)}\right]^{x_n}\,. \ee Clearly this cannot make sense when
the argument of the power-law becomes negative (i.e. for
$t>2\ell$), signaling that there is something wrong in the
derivation. However, using the replica trick, for the entanglement
entropy we obtain 
\be
S_A=\frac{c}{6}\log\left(\frac{4\ell}{\pi\e}t\sin(\pi
t/2\ell)\right)+\tilde{c}'_1\,. 
\label{S2} \ee 
One is tempted to
assume that this result can be correct only for $t<\ell$ and that
for larger time it saturates as suggested by the quasiparticle
interpretation. Only an exact calculation via the exact mapping
(\ref{exmap}) can resolve these doubts. However exact solutions of
integrable models or numerical density matrix renormalization
group could already be able to eventually exclude Eq. (\ref{S2})
and to shed some light on the problem. We mainly mention this
topic here to encourage further studies in this direction.

Finally let us point out that transformations similar to Eq. (\ref{exmap})
appear when studying finite-size effects.

\section{Discussion}
\label{secC}

We presented a detailed study of the unitary evolution that
results after a local quench in a quantum one-dimensional system.
All our findings have been obtained by means of CFT and so they
are limited to the evolution of a gapless quantum model with a
linear dispersion relation. We presented calculations for the
entanglement entropies for different bipartitions of the system
and for one- and two-point correlation functions in a general CFT.
In particular we studied the case when two half-systems are joined
at time $t=0$.

All our results are interpretable in terms of a scenario that we
believe to be valid in general, not only for gapless systems (in
analogy to the case of global quenches \cite{cc-05,cc-06,cc-07}).
In fact, the initial state is expected to generate quasiparticle
excitations at $r_D=0$ that then propagate freely through the
system and carry all the information about entanglement and
correlations. In the case of a CFT all these excitations travel at
the same speed $v_s$ ($=1$ by normalization). However in general,
as a consequence of a non-linear dispersion relation, a full
spectrum of velocities is expected.

Although the physical cut may seem a rather specific situation, we
believe that our findings are quite general. For example, a
general defect (e.g. a weakened bond) is asymptotically equivalent
to the results we obtained here as long as the defect is a
relevant perturbation. This excludes the XX chain that is the only
model studied so far \cite{ep-07},  where the defect is believed
to be marginal \cite{p-def}, but should apply to XXZ spin-chains
\cite{def2}. Furthermore we expect a qualitatively similar
behavior when a system has been artificially prepared in a
configuration that is only locally different from the ground
state. For example there are already several studies where the
local horizon effect is evident (see e.g. those considered in
Refs. \cite{lochor}). A full quantitative analysis of most of
these settings should be possible properly adapting our CFT
treatment (e.g. with the theory of boundary condition changing
operators \cite{al-95}).
On the other hand, also the case of an initial state with one or more kinks 
displays a time-evolution with an horizon effect similar to the one
considered here (see e.g. Refs. \cite{kink} where also finite temperature
results are obtained in a rigorous manner). 
It would be interesting to understand whether these results can be recovered
analytically continuing some CFT results. 
Work in this direction is in progress.

We described here several different physical situations, but many
problems are still left for future investigations. A first goal
would be to check our predictions in exactly solvable models. In
fact, as far as we are aware only the paper by Eisler and Peschel
\cite{ep-07} discussed these topics in the context of the XX model
and their results are fully compatible with ours.

Another simple generalization of our results is to study gapped exactly
solvable models, like the Ising model in a transverse magnetic field or others
admitting a free-field representation. In these case it is also interesting to
understand whether some predictions can be made on the basis of the generalized
Gibbs ensemble \cite{gg}.

A more difficult question to study analytically concerns the role
played by quenched disorder. This is expected to change at a
qualitative level the quasiparticle scenario as a consequence of
Anderson localization \cite{bo-07}. In this case the
time-dependent density matrix renormalization group \cite{tdmrg}
should be quite effective, since for clean systems we know that
the entanglement entropy never increases dramatically as it does
in the case of global quenches.

\section*{Acknowledgments}
This work was supported in part by EPSRC grants GR/R83712/01 and
EP/D050952/1. This work has been done in part when PC was a guest of
the Institute for Theoretical Physics of the Universiteit van
Amsterdam. This stay was supported by the ESF Exchange Grant 1311 of
the INSTANS activity. PC thanks Dragi Karevski for useful discussions.

\end{document}